# Visualizing coexisting surface states in the weak and crystalline topological insulator Bi$_2$TeI

Nurit Avraham[1], Andrew Norris[1], Yan Sun[2], Yanpeng Qi[2], Lin Pan[2], Anna Isaeva[3], Alexander Zeugner[3], Claudia Felser[2], Binghai Yan[1] and Haim Beidenkopf[1]

[1]Department of Condensed Matter Physics, Weizmann Institute of Science, Rehovot 7610001, Israel
[2]Max Planck Institute for Chemical Physics of Solids, D-01187 Dresden, Germany.
[3]Technische Universität Dresden, Bergstraße, 66, Dresden, D-01069, Germany.

**The established diversity of electronic topology classes lends the opportunity to pair them into dual topological complexes. Bulk-surface correspondence then ensures the coexistence of a combination of boundary states that cannot be realized but only at the various surfaces of such a dual topological material. We show that the layered compound Bi$_2$TeI realizes a dual topological insulator. It exhibits band inversions at two time reversal symmetry points of the bulk band which classify it as a weak topological insulator with metallic states on its (010) 'side' surfaces. Additional mirror symmetry of the crystal structure concurrently classifies it as a topological crystalline insulator. Bi$_2$TeI is therefore predicted to host a pair of Dirac cones protected by time reversal symmetry on its 'side' surfaces and three pairs of Dirac cones protected by mirror symmetry on its 'top' and 'bottom' (001) surfaces. We spectroscopically map the top cleaved surface of Bi$_2$TeI, and crystallographic step edges therein. We show the existence of both two dimensional surface states which are susceptible to mirror symmetry breaking, as well as one dimensional channels that reside along the step edges. Their mutual coexistence on the step edge where both facets join is facilitated by momentum and energy segregation. Our observations of a dual topological insulator make way to additional pairing of other dual topology classes with distinct surface manifestations coexisting at their boundaries.**

# INTRODUCTION

Band insulators are classified into different classes according to the topology of their band structures [1-4]. In the presence of time reversal symmetry, insulators are divided into trivial insulators and topological ones, with the latter hosting symmetry-protected metallic states on their surfaces. Lattice symmetries further refine the classification and introduce subclasses into the trivial class, for which only some of the surfaces carry topologically protected metallic states. Examples include the weak topological insulator (WTI) [4] for which the protecting symmetries are time reversal symmetry and a particular discrete translational invariance; and crystalline topological insulators (TCI) [5,6] for which the protecting symmetries are crystalline symmetries such as mirror and rotation. Remarkably, the same material can host surface states of different nature on different material facets that originate from the same bulk band inversions. Here we provide experimental realization of such dual TI that exhibits different surface states on different facets.

Dual topological classification can take different forms of realization. One form may refer to dual protection of surface states by multiple bulk symmetries. This is the case in the canonical strong topological insulators $Bi_2Se_3$ and $Bi_2Te_3$ [7] and in other compounds [8]. On top of time reversal symmetry these compounds host three mirror planes related by C3 symmetry which simultaneously classify them as topological crystalline insulators. Consequently, in the presence of magnetic field in a direction that preserves the mirror symmetry the Dirac surface states would not gap out but merely shift from the time reversal invariant momentum at Γ. A different type of dual topological classification arises when distinct topological indices can be assigned at different momenta in the Brillion zone in a given material. Consequently, topological surface states of distinct types will coexist on the same surface. For instance, Dirac and Fermi-arc states were predicted to coexist both in LaPtBi [9] and in $Cu_2S$ [10] due to their dual bulk classification as Weyl semimetals and strong topological insulators. A different realization of such coexistence was shown to be stable at the interface between a Weyl semimetal and a strong topological insulator [11]. A third kind of dual topological classification, which stands in the focus of this study, occurs when multiple topological indices can be assigned to the bulk band structure, however these give rise to different topological surface states on different surfaces of a given material. This was predicted to occur in $Bi_2TeI$ [12], which we study here, and in $Bi_1Te_1$, in which only the existence of the TCI surface states was directly imaged and not their dual topological nature [13]. In both compounds, an even number of bulk band inversions can be simultaneously assigned with weak topological indices and mirror Chern numbers. Consequently, time reversal symmetry protected Dirac states are expected to form on one set of surfaces and mirror symmetry protected Dirac states will form on a distinct set of surfaces. Intriguingly, these mutually existing surface states coexist along various hinges in the material which, as we show here, allows them to interact with one another.

The layered compound $Bi_2TeI$ [12,14-16] (Fig.1A) is composed of stacked layers in which a Bi-bilayer is intercalated between two layers of the trivial semiconductor BiTeI (Fig.1B). A Bi-bilayer was shown to be a two dimensional topological insulator (2DTI) [17,18], hence a unit cell layer of TeBiI-$Bi_2$-IBiTe is expected to be a robust 2DTI [15]. In general, when layers of 2DTIs are stacked and weakly coupled, they form a WTI [4, 19]. Recent studies have predicted that a mirror protected TCI can be also described using a coupled layer construction [20]. Indeed, theoretically, 3D $Bi_2TeI$ was predicted to have a dual topological classification inclusive of both a WTI and a TCI [12]. More recently it was predicted that when some of its symmetries are broken $Bi_2TeI$ is classified as a higher order topological insulator, which exhibits topological hinge modes [21-23].

The dual WTI-TCI classification can be derived from the two band inversions located at the $\Gamma$ and Z time-reversal symmetric points of the 3D bulk Brillouin zone (Fig.1C) [12,16]. Projecting these inverted bulk bands onto the (010) 'side' surfaces, two surface Dirac cones form - one at the $\tilde{\Gamma}$ point and the other at the $\tilde{Z}$ point of the (010) surface Brillouin zone [15]. On the (001) 'top' and 'bottom' surfaces, the two inverted bulk bands are projected to the same $\overline{\Gamma}$ point. The two surface Dirac bands should therefore hybridize and fully gap out, leading to the characteristic WTI surface phenomenology comprising of metallic side surfaces and insulating top and bottom surfaces. However, the existence of three $\Gamma$-Z-M mirror planes in $Bi_2TeI$ (Fig. 1A) protects the band crossing points along their mirror-projection lines ($\overline{\Gamma} - \overline{M}$) on the (001) surface [12]. In fact, two of the three mirror planes are broken by a slight distortion [12]. The induced gap, however, is below our resolution, hence we neglect here this distortion. As a consequence of those mirror symmetries, three pairs of Dirac surface states are expected on the (001) surface band structure (Fig.1E). This classifies $Bi_2TeI$ as a TCI together with its classification as a WTI. We study the intricate coexistence of the WTI and TCI surface states (Fig.1F). In particular, we focus on facet edges where the two types of states interact and examine their interplay as well as their individual and mutual response to disorder. By mapping the local density of states (LDOS) on terraces and step edges on the (001) surface, we directly visualize, 1D metallic channels coexisting alongside two-dimensional topological surface states. The TCI surface states appear on all surface terminations and are susceptible to the mirror symmetry breaking imposed by the step-edges. The 1D channels appear only on step edges which contain a Bi-bilayer, and are absent at step edges that lack that topological element. The two types of surface states can coexist due to the large momentum transfer involved in their hybridization.

# RESULTS

In the following, we present dI/dV(V) measurements on the top (001) surface of $Bi_2TeI$. In the first section we visualize the TCI states and demonstrate their susceptibility to mirror symmetry breaking perturbation. Next, we visualize 1D channels along step edges and demonstrate their topological nature by their selective appearance at Bi-bilayer containing step edges only. To that end we carefully characterize the atomic structure of the various terraces that we probe. Finally, we study the interplay between the TCI surface states and the 1D edge channels. We show that while in relatively clean regions they coexist due to separation in energy and momentum, such protection cannot withstand strong disorder and leads to partial hybridization between the two types of states.

**TCI surface states under mirror symmetry breaking perturbation**

Single crystals of $Bi_2TeI$ (See materials and methods in the supplementary) were cleaved in ultra-high vacuum conditions and measured in a scanning tunneling microscope (Unisoku). Topographic and differential conductance (dI/dV) measurements of the LDOS on the sample surface were then carried out at 4.2K. Cleaving the sample in commensuration with its layer orientation exposes a fresh surface along the (001) direction with multiple terraces and islands of different heights. Fig. 2A shows a typical topographic image of such a cleaved surface, exhibiting a series of several adjacent terraces that end at rough crystallographic step edges (right inset). Although the easy cleavage plane is along the two van der Waals coupled Tellurium terminations

[15], we find that it actually occurs also along other van der Waals coupled layered, hence different terraces can have different surface termination. As we show later on, we uniquely identify the terraces layer ordering and surface terminations based on their step heights, spectral and structural characteristics.

To directly observe the TCI surface states we measured the LDOS on the surface of the different terraces. A typical dI/dV(V) spectrum measured on a Bi-terminated surface (at the location marked by a blue arrow in Fig. 2A) is given by the solid line in Fig.2B. The surface spectrum shows metallic behavior with reproducible peak-like features at several characteristic energies. We compare the measured spectrum to *ab initio* calculations of the bulk and surface density of states (see supplementary for detailed *ab initio* calculations). Good agreement of the measured peak-like features with those found in the calculated bulk density of states (greyed area in Fig. 1B) reveals that these correspond to bulk bands. However, in contrast to the calculated semiconducting bulk spectrum with a bulk gap of about 160 meV the spectrum we measure on the surface is fully metallic. This apparent discrepancy is resolved by calculating the surface density of states projected on the Bi-terminated (001) surface (dotted line in Fig. 2B), which shows excellent agreement with the measured curve throughout the spectrum. The in-gap surface LDOS is therefore entirely contributed by surface states. In the calculation these correspond to the TCI Dirac surface states with the Dirac node at the vicinity of the Fermi energy regardless of surface termination (for more details see supplementary Fig.4).

Below the Fermi energy, we observe signatures of dispersing LDOS modulations (Fig. 2B inset, left panel) indicative of quasi particle interference (QPI) due to scattering off the step edge. Indeed, the Fourier transform of this dI/dV map (inset, right panel) shows the dispersion of these QPI patterns up to an energy of about -50 mev in agreement with the calculated onset of the bulk semiconducting gap. Within the bulk gap, where the TCI states reside, clear QPI patterns are no longer observed (inset, right panel and supplementary). Helical Dirac surface states are indeed known to give rise to a rather faint QPI signatures due to the suppression of the backscattering channel, which is otherwise the strongest. In addition, we find that about 5% of the surface atomic sites show a height variation whose extent seems to correspond to an adatom rather than a substitutional defect. Such mean adatom separation might be too short compared with the electronic wavelength to clearly resolve the expected long wavelength QPI pattern of the Dirac surface bands. The rough contour of the step-edges further limits the ability to resolve surface state QPI. We note, that the observed cleaved surface morphology resembles that of the cleaved surface of the trivial semi-conductor BiTeI [24]

Lacking a direct visualization of the metallic in-gap surface states dispersion leaves their identification with topological TCI states rather than trivial surface bands ambiguous. A distinction among them can be achieved by applying a symmetry breaking perturbation that selectively affects only the topological surface states [6, 25]. For the TCI surface states in $Bi_2TeI$, such a symmetry-breaking perturbation is provided by the step-edges that naturally appear on a cleaved (001) surface and break the mirror symmetry of the lattice. We therefore measured the dI/dV(V) spectrum (Fig. 2C) along a line cut (grey line in Fig. 2A) taken across all seven terraces. We find that *on* the terraces, far from the edges, the spectrum is predominantly metallic irrespective of the specific cleaved termination. Such an unselective metallic behavior is uncommon to non-topological Shockley or dangling-bond surface states whose existence is strongly surface dependent [12, 15]. It is characteristic of the TCI surface states, that are guaranteed by the bulk topology, and are therefore expected to exist on all (001) surface terminations.

Remarkably, in the face of this robustness, we find a consistent suppression in the LDOS of the in-gap surface states at the vicinity of all six crystallographic step edges (colored in pale blue). The spatial extent of the suppressed LDOS is over about 20 nm. Representative density of states spectra taken near and far from a step-edge (at the blue and black arrows in Fig.2C, respectively) are directly compared in Fig. 2D. This clearly demonstrates the suppressed in-gap LDOS over about 50 meV above and below the Fermi energy, in the vicinity of the step-edge. Trivial surface states, as well as bulk states including impurity bands, are hardly susceptible to such crystal mirror symmetry breaking. The typical response of such surface states to crystallographic step-edges is in the form of scattering which does not induce any gaps, certainly not over such extended length scale. Impurity bands, in which translational invariance is altogether broken, would be even less susceptible to such disorder. TCI surface states, on the other hand, are highly susceptible to mirror symmetry breaking, and would be affected over a length scale governed by their wavelength, that diverges towards the Dirac node. Accordingly, the observed LDOS suppression identifies the in-gap metallic surface states we measured with the mirror-symmetry protected TCI surface states expected on the (001) surface of $Bi_2TeI$ and the energy of their Dirac nodes to be within the 100 meV energy range in which the LDOS got suppressed.

We find the in-gap LDOS to decrease at the vicinity of the step edges but not to vanish. This is indeed the expected behavior of the mirror protected topological surface states considering their three protecting mirror planes. A straight crystallographic step-edge that appears on the (001) surface breaks at least two out of the three mirror symmetry planes of the crystal. As a result, out of the six Dirac cones that reside on the surface, the ones that were protected by those mirror planes will gap out. This is shown schematically at the inset of Fig.2D. Moreover, the contours of the step-edges we observe on the surface of $Bi_2TeI$, shown in Fig.2A and at its right inset, are not atomically straight. The LDOS of the TCI states is therefore expected to decrease at the vicinity of all step-edges, but the extent of this suppression is expected to change with the local morphology of the step-edge. The remaining in-gap LDOS is accordingly contributed by the ungapped TCI states, and possibly by non-topological surface bands that may coexist on some of the surfaces [9], or due to impurity states. An important observation is that for all step-edges shown in Fig. 2C, the LDOS is significantly suppressed only on the upper side of the terrace (demonstrated at the inset). We attribute this asymmetry to the asymmetric boundary condition a step-edge imposes on the surface states in the upper and lower terraces. For the wave function on the lower terrace, the layer in which it resides continues through the step-edge, while for the wave function in the upper terrace the layer terminates abruptly at the step-edge. As a result, the gaping induced by the mirror symmetry breaking is expected to be stronger on the upper side of the step-edge. This, again, signifies the susceptibility of the imaged surface states to the extent of mirror symmetry breaking imposed by crystallographic defects and alludes to their topological nature.

## 1D edge channels of a WTI

Having verified the existence of TCI states on the terraces' surface and visualized their response to mirror symmetry breaking, we now turn to investigate the properties of the WTI states. Unlike the TCI states that showed insensitivity to the 'top' terrace termination, the composition of the 'side' surface, which during cleave gets fragmented into well separated step-edges, is detrimental to the formation of topological boundary states. Since the Bi-bilayer is the essential building block of the WTI surface states, a step-edge of a terrace terminating a single Bi-bilayer in its layer composition is expected to host a 1D conducting helical channel that is protected by time reversal symmetry [26, 27, 28]. Similar 1D channels were measured previously on the surface of the WTI $Bi_{14}Rh_3I_9$ [29, 30]. While in generic WTIs these channels appear on the background of a gapped surface, in $Bi_2TeI$ they would coexist with the TCI surface states, lending a unique opportunity to investigate their interaction.

In Fig. 3 we show visualization of 1D conducting channels measured on step-edges on the surface of two different samples of $Bi_2TeI$. The crystallographic step between terraces 2 and 3 (presented in Fig.2A) has a unit cell height of 1.7 nm. In contrast to smaller steps that may lack Bi-bilayers, or to higher steps that may host two coupled Bi-bilayers, a step of a unit cell height must contain only a single layer of Bi-bilayer. As a result, a 1D topological channel is expected at its boundary. The locally averaged dI/dV(V) curves measured right at this step-edge along with the ones measured on top of the two adjacent terraces are shown in Fig.3A (solid and dashed lines, respectively). The two different terraces exhibit almost overlapping dI/dV(V) curves with a finite in-gap LDOS contributed by the TCI surface states. The dI/dV(V) measured on the step-edge is clearly distinguished. It exhibits an additional significant increase in the in-gap LDOS that peaks 100 meV (red arrow) below the onset of the conduction band (black arrow) and extends across the bulk gap. We note that this peak energy is well below the typical energy window over which the TCI surface states are suppressed at the step-edges, allowing the simultaneous observation and identification of the appearance of the WTI 1D channels and the suppression of the TCI states. Our DFT calculations (supplementary Fig. 5) show similar peak-like structure of a 1D topological channel residing at a step-edge crystal termination. Spectroscopic maps measured at different locations across step-edge 2-3 (Fig. 3B) clearly visualizes a 0.5 nm wide 1D channel running along the step-edge. Similar spectroscopy of a 1D edge channel was measured at step edges 5-6 (Fig. 3C), as well as on a unit cell high step-edge measured on a different sample (Fig. 3D). In the other sample, however, the Fermi energy resides at the top of the valence band rather than at the bottom of the conduction band. Yet, in both samples the characteristic peak-like feature appears 100 meV below the onset of the conduction band.

One dimensional channels at step-edges can result also from non-topological origins such as dangling bonds at the terrace boundaries. While the existence of these channels is independent of the terace's Bi bilayer content, the formation of the topological channels must be correlated with the existence of a Bi bilayer in the terrace layer structure. Terraces containing a single or an odd number of Bi-bilayers must host topological channels, while step-edges with no Bi-bilayers are not expected to host topological channels. In the case of an even number of Bi-bilayers, the edge states may gap out in pairs [4]. We examined the relation between the formation of 1D channels and the Bi bilayer content of the measured step-edges. To determine the Bi-bilayer content we resolved the layer termination of the whole staircase structure. As shown in Fig.3E, by correlating the staircase height profile with the layer structure of $Bi_2TeI$, we find that there is only one configuration for which all terraces cleave planes fall between Van der Waals coupled layers. All other configurations (see supplementary Fig. 6) involve breaking some covalently bonded layers which makes them improbable. We further verified that favorable configuration by cross correlation analysis of the atomically resolved topographic images measured on the various terraces (See supplementary Fig. 7), as well as by the terraces spectral characteristic. Crosschecking the information of these independent analysis methods allowed us to uniquely determine the layer ordering of the whole staircase (Fig.3E). We find that terraces 2, 3 and 6 are terminated by Bi-bilayers while terraces 4, and 5 are terminated by Te and I, respectively. Comparing this identification with the measured dI/dV(V) shows excellent agreement. Step-edges *2-3,* and *5-*6, *whose upper terraces are terminated by a Bi bilayer*, indeed host topological channels (Fig.3A-3C). However, step-edges *3-4* and *4-5, with no Bi Bilayer content,* do not show any increase in the in-gap LDOS with respect to the spectra measured on their adjacent terraces, nor the characteristic peak that we consistently identify with the topological 1D channel (Fig. 3F and 3G respectively). We thus confirm complete correlation between the presence or absence of 1D metallic edge channel at the step-edge and the presence or absence of a Bi-bilayer in the step-edge layer composition. This suggests the topological nature of the observed 1D channels. It also rules out the identification of the 1D channels with TCI 1D edge modes that were recently suggested to exist in SnTe [31], since these should also be termination independent. Furthermore, the existence of those 2D states does not undermine the existence of the 1D metallic channels signifying their fair level of decoupling in the dual topological insulator.

## Interaction between the TCI and WTI surface states coexisting on hinges

A striking observation in Fig.3B is that the 1D edge channels coexist with the adjacent 2D TCI surface states, rather than hybridizing with them. We attribute this decoupling to the large momentum transfer required for their hybridization [32, 33]. This is demonstrated in Fig.1F by the combined band structure of the 2D TCI states and the 1D edge channel. The mirror protected Dirac cones are located on the $\bar{\Gamma} - \bar{M}$ directions away from the $\bar{\Gamma}$ point [12]. In contrast, the 1D state, which is contributed by the Bi-bilayer [12,15] and is protected by time reversal symmetry, is bound to the time reversal symmetric $\bar{\Gamma}$ point (Fig.1B). Hence, the hybridization of the two is suppressed by the large momentum separation and they are topologically protected from gapping in a manner similar to that exhibited by bulk Weyl states of a Weyl semimetal [34, 35]. Nevertheless, the level of protection is somewhat diminished, as it depends on the ability of scatterers to provide the finite momentum transfer. Accordingly, the ability to resolve these 1D channels as well as their spatial distribution along the step-edge, is strongly affected by the high level of disorder observed at the step-edges (insets in Fig.2A).

We indeed find that at some disordered regions the channels appear quite fragmented as opposed to the continuous channel imaged on the straight segments of the step edge in Fig.3B. We map the spatial distribution of the wave-function of the edge channel on the disordered terrace next to the rough step-edge whose topography is shown in Fig.4A. The complex topographic landscapes imaged in Figs.3A and 4A (and supplementary Figs. 6 and 7) results from the combination of resolved crystal structure, disorder and the spatially modulated LDOS response to such disorder. Indeed, the dI/dV(V) spectra in Fig.4B, taken at different positions across the terrace, show a varying intensity of the -100 meV peak characteristic of the edge channel. We Spatially map the intensity distribution of the edge channel wavefunction across the terrace in Fig.4C. Remarkably, we find that while in some segments it is bound to the step edge, in others it becomes fragmented. This behavior stands in sharp contrast to the robustness demonstrated for the edge modes both in the WTI $Bi_{14}Rh_3I_9$ [29] and in Bi-bilayes [17]. The imaged wavefunction distribution bears resemblance to previously imaged localized states [36,37], though these differ in dimensionality and their topological nature. Chemical disorder is thought to lead to fluctuations in the spatial distribution of the wave function of the 1D channels. While these channels are expected to be immune to spatial disorder, in order to avoid it they might penetrate deeper into the bulk or sideways, away from the step-edge and into the terrace surface. When mapped locally, this would appear as their local disappearance or their spatial fragmentation. Still, this fragmented structure seems to be unique to the hinges in $Bi_2TeI$ where the two distinct surface states of the WTI and TCI coexist.

# SUMMARY

We verified the existence and visualized the coexistence of TCI and WTI surface states in $Bi_2TeI$. The TCI Dirac surface states appear on all terraces, regardless of composition and termination, and are found to be susceptible to mirror symmetry breaking by the step edges. The WTI edge modes, on the other hand, appear selectively only on those step edges at which a Bi-bilayer terminates. Their coexistence is enabled by these surface states being separated in real space, momentum space and energy. In real space the separation is a consequence of the different facets on which the TCI and WTI surface states reside, which still allows their interaction on facet edges. In momentum space it originates from the confinement of the 2D TI Dirac node to time reversal invariant momentum, while no such confinement occurs for the TCI. On the energy axis the separation is a result of the details of the energy dispersion. Our measurements highlight the unique interplay of these protection mechanisms. We observe the states on terraces and step-edges where their spatial separation is minimal. Remarkably, they remain decoupled due to their separation in momentum space and in energy. On a practical point of view our findings raise the possibility of engineered samples with designed terraces, step edges and symmetry breaking perturbations. Such devices will have defined regions whose properties are determined by the specific combination of topological surface states they host.

## MATERIALS AND METHODS

$Bi_2TeI$ single crystals were grown via self-flux method using Bi, $I_3$, Te and Bi as starting materials. A mixture with composition $Bi_2TeI$ was placed on alumina crucible sealed in an evacuated quartz tube. The mixture was heated at 300°C for 10 h and 550°C for 20 h, then slowly cooled to room temperature at a rate of 2°C/h. The obtained crystals were silver-gray with a typical dimension of 2x2x0.2 $mm^3$. The $Bi_2TeI$ single crystal where cleaved in the STM load lock at UHV conditions, at room temperature, using the scotch tape technique. All dI/dV(V) measurements were taken at

4.2 K using standard locking techniques. Typical parameters are lock-in frequency of 733 Hz, ac amplitude of 5 mV, parking bias of -300 meV, and current set point of 300 pA.

## Acknowledgements

N. A., H. B., B. Y. and C. F. acknowledge the German-Israeli Foundation for Scientific Research and Development (GIF Grant no. I-1364-303.7/2016). H. B. and N. A. acknowledge the European Research Council (ERC, Project No. TOPO NW), B. Y. acknowledges the support by the Ruth and Herman Albert Scholars Program for New Scientists in Weizmann Institute of Science, Israel. C. F.'s work was financially supported by the ERC Advanced (Project No. 742068, 'TOPMAT'). We are grateful to Ady Stern for fruitful discussions and to Hadar Eizenshtat for his contribution to the measurements.

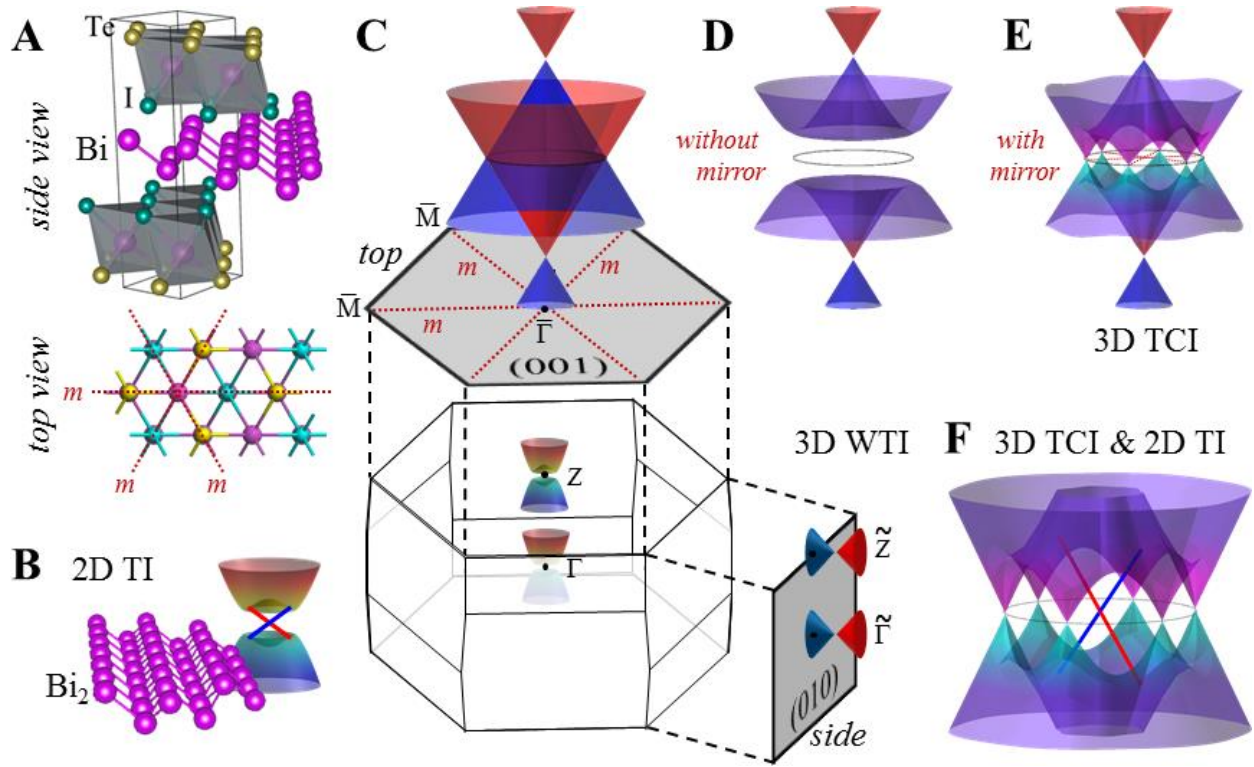

**Figure 1. Dual topological classification in Bi$_2$TeI**
(A) The side view of the Bi$_2$TeI layer structure shows layers of the 2DTI Bi-bilayer sandwiched between layers of the non-topological semiconductor BiTeI. The top view captures the three mirror symmetry axes of the trigonal structure of the layers. We neglect here the slight distortion that reduces the trigonal structure of the layers down to a monoclinic symmetry of the entire Bi$_2$TeI. (B) A single Bi-bilayer is a 2DTI with helical metallic channels running on its edges. (C) Two inverted gaps located at the Γ and Z high symmetry points of the 3D bulk BZ. Their projection on the side (010) surface yields two Dirac cones at $\tilde{\Gamma}$ and $\tilde{Z}$ that forms a WTI. Their projection on the top (001) surface yields two overlapping Dirac cones that are slightly shifted in energy, forming a degenerate nodal line. (D) *Without mirror-symmetry* the Dirac cones hybridize and the nodal line may gap out (E) *With mirror-symmetry* the nodal-line is protected on the $\overline{\Gamma} - \overline{M}$ mirror-projection lines, leading to six Dirac cones on the top surface. (F) The combined surface spectrum of the (001) surface with 1D channel along crystallographic step-edges consists of six 2D Dirac cones along the $\overline{\Gamma} - \overline{M}$ directions and a helical 1D state at the $\overline{\Gamma}$ point.

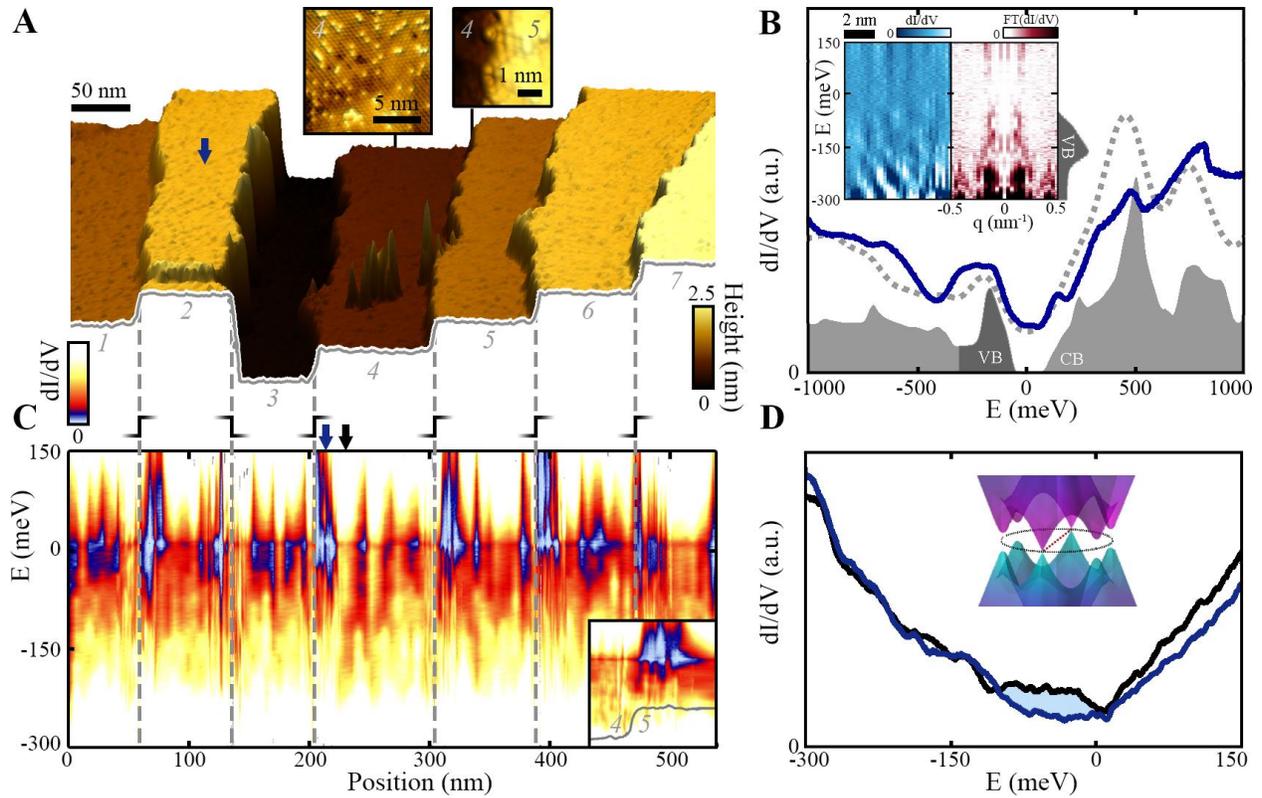

**Figure 2. TCI surface states under mirror symmetry breaking perturbation.**
**(A)** Topographic image of a staircase structure on the surface of $Bi_2TeI$. Only step-edge 2-3 is unit cell high, the rest are smaller. The left inset shows typical point-like disorder. The right inset shows typical atomic scale roughness of the step edges. **(B)** Comparison of the normalized LDOS (solid line) measured at the blue arrow in A, and the calculated bulk and surface-projected density of states (dotted and gray shaded, respectively). The inset shows QPI patterns in the LDOS along with their Fourier transform. These QPI decay when approaching the energy gap **(C)** False color plot of the LDOS measured along the gray line profile in A showing metallic behavior throughout the surface, with partial suppression in density of states next to the mirror symmetry breaking step edges. The inset shows the asymmetry between the suppression on the top terrace of a step edge (corresponding topographic profile overlaid) and its absence on the bottom terrace. **(D)** The amount of suppression of the in-gap states (blue shaded region) next to a step-edge versus far from it (blue and black lines, respectively)

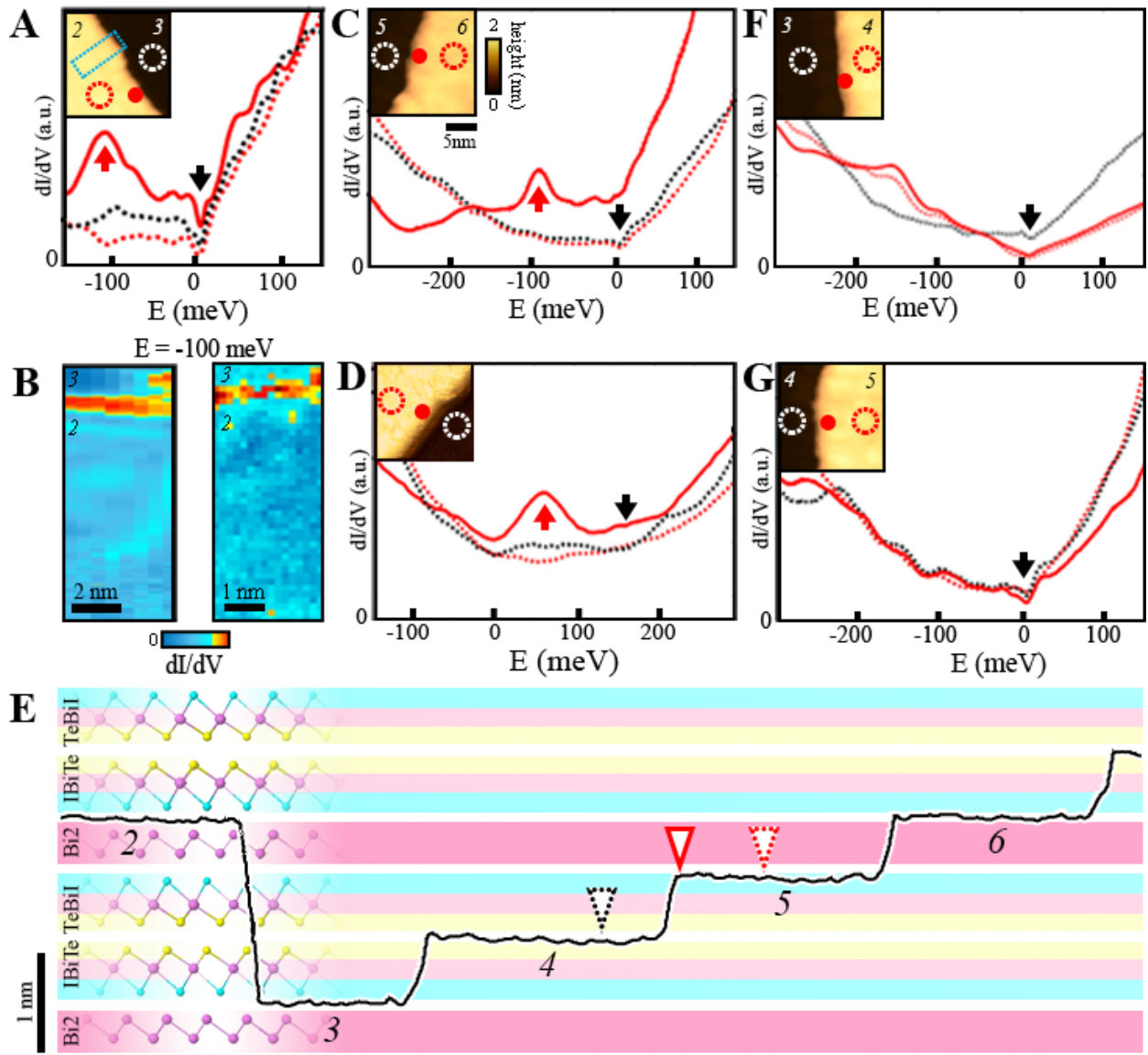

**Figure 3. 1D edge channels of a WTI in the presence of disorder**
(A) Step-edge 2-3 in Fig.2A: dI/dV(V) spectra locally averaged over a few squared nm regions located close to the step-edge termination and slightly away from it on both sides of the step-edge (solid and dotted lines measured at the locations marked by the solid and dotted tips in A), the 1D channels are characterized by a peak like feature at -100 meV. Inset:topgraphy of the step edge. (B) dI/dV(V) map of the 1D edge channel taken at its energy of peak intensity of -100 meV at two different regions. (C) Same measurements as in A for step-edge 5-6. (D) Same as in A for a different sample. Other than a 160 meV shift of the fermi energy, the spectroscopic features including the peak-like features are identical to those in A and C. (E) Height profile of the staircase structure (same as in Fig.1A) overplayed on the layer structure of Bi$_2$TeI indicated by the schematic side view of the crystal structure. (E) and (G) dI/dV(V) point spectra as in A for step edges 3-4 and 4-5, respectively, where no 1D channels where observed.

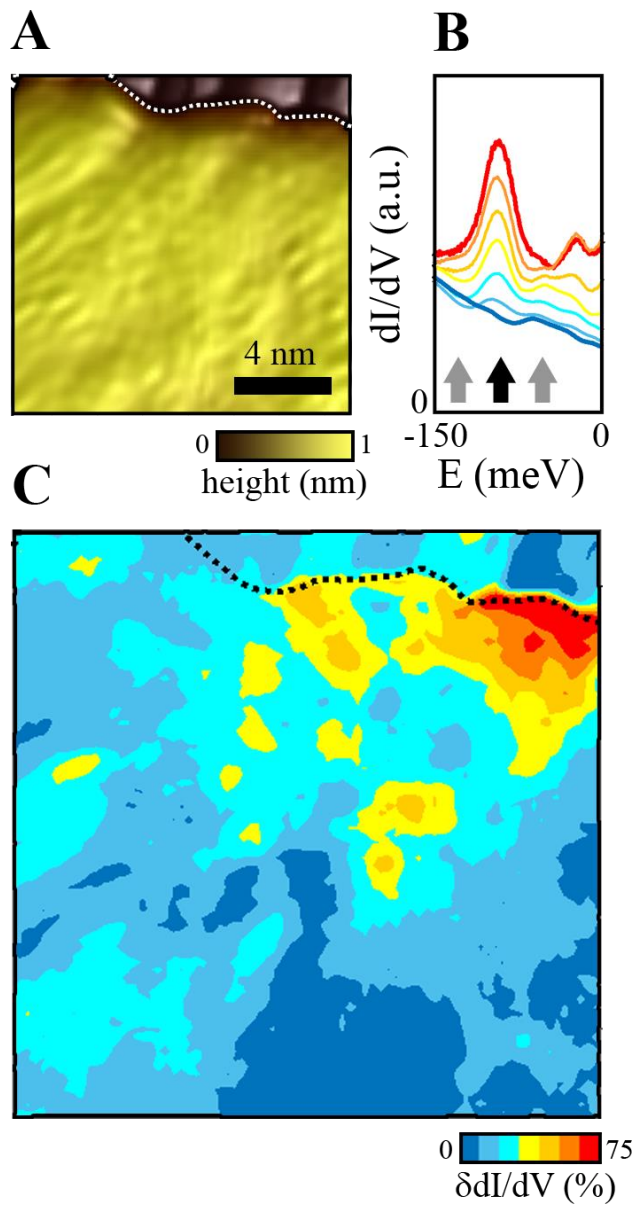

**Figure 4. Interaction between the TCI and WTI surface states coexisting on hinges**

(A) Topographic image of a rough segment of step edge 5-6 (edge contour highlighted with dotted line) terminating a disordered terrace as seen by its fluctuating height. (B) dI/dV(V) point spectra averaged over different regions showing varying resonance intensity at -100meV relative to the background density of states (black versus gray arrows, respectively). (C) Mapping of the relative -100meV resonance intensity (amount of change between dI/dV(V) value at black and gray arrows) shows how the edge channel (edge contour shown by dotted line) gets fragmented in face of the edge roughness and surface disorder.